\documentclass[a4paper,12pt]{article}
\usepackage[left=1in,right=1in,top=1in,bottom=1in]{geometry}
\usepackage[utf8]{inputenc}
\usepackage{epsfig}

\usepackage{float}
\usepackage{xcolor}
\usepackage{multirow}

\def\R{{\cal R}}
\def\H{{\cal H}}
\def\C{{\cal C}}

\title{\textbf{Bargmann Invariants, Geometric Phases and Recursive Parametrization with Majorana Fermions}}

\author{Rohan Pramanick\footnote{rohanpramanick25@gmail.com} , Swarup Sangiri and Utpal Sarkar\\
\small{Department of Physics, Indian Institute of Technology Kharagpur, Kharagpur 721302, India}
}
\date{}

\begin{document}
	

\maketitle
\thispagestyle{empty}

\begin{abstract}
A generalized connection between the quantum mechanical Bargmann
 invariants and the geometric phases was established for
 the Dirac fermions. We extend that formalism for the Majorana
 fermions by defining
 proper quantum mechanical ray and Hilbert spaces. 
 We then relate both the Dirac and Majorana type Bargmann invariants to the rephasing 
 invariant measures of CP violation with the Majorana 
 neutrinos, assuming that the neutrinos have lepton number violating Majorana masses. 
 We then generalize the recursive parametrization for studying any unitary
 matrices to include the Majorana fermions, which could be useful for studying the neutrino mixing matrix. 
\end{abstract}


\section{Introduction}

One of the most interesting problems of the standard model is to
understand the origin of CP violation. It appears in
different forms and was first observed in weak decays of the 
neutral K-mesons. CP violation is also needed
to explain why there are more matter compared to antimatter
in the universe \cite{sakh}. 

CP violation has been studied extensively for 
the Dirac fermions \cite{jarlskog,green}. 
All known charged quarks and leptons are Dirac particles and 
their analysis does not have direct implications to the 
lepton or baryon number violating interactions, including
the generation of matter asymmetry of the universe.
We thus attempt to generalize some results for the Dirac
fermions to models with Majorana fermions like the neutrinos. 

Although we
are yet to infer if there exists any Majorana fermion, many interesting
aspects of the Majorana fermions have been pointed out, which
may have far reaching consequences. In particle physics, the masses
of the Majorana particles play a crucial role and can explain
the smallness of the neutrino mass naturally \cite{seesaw}. The Majorana neutrino
masses can also explain the baryon asymmetry of the universe \cite{fy} 
and predict the dark
matter and resolve some issues of the dark energy. 

Considering all these, we intend to study the CP phases for
the Majorana fermions from a different angle. There have been 
some analysis of the Dirac fermions (quarks and leptons \cite{ql}) identifying their rephasing invariant measure of quantum 
mechanical CP phases to the quantum mechanical Bargmann 
invariants (BI) \cite{bargmann}, which in turn, is related to the 
classical geometric phases. 
Our main aim is to generalize this result for the Majorana 
fermions and apply our result to the CP violation in the leptonic
sector with Majorana neutrinos. 

The geometric phase was introduced \cite{berry} in a cyclic adiabatic quantum
mechanical system, where the dynamics is governed by the time-dependent 
evolution of the 
state vector in a Hilbert space. These geometric phases has been 
shown to be related to a family of
quantum mechanical Bargmann invariants (BI). For a
physical system, the state vectors represent Dirac fermions, and
the BIs may be identified with rephasing invariant measures of CP
violation. We shall generalize these results to the case when the state
vectors represent both Dirac and Majorana fermions by defining the 
ray and the Hilbert spaces properly. This will introduce additional
BIs representing CP violation arising from the Majorana phases and 
this, in turn, will relate 
the rephasing invariant measures of CP violation for both Dirac 
and Majorana fermions to the complete sets of Bargmann invariants.

We shall first demonstrate how one can define the quantum
mechanical Bargmann variables for the Majorana fermions and relate
them with the geometric phases after defining the proper quantum
mechanical ray and Hilbert spaces for the Majorana fermions. 
We then construct the rephasing invariant measures for the Majorana
fermions. As expected, compared to the Dirac fermions, there are
more number of such CP violating measures arising due to the Majorana 
phases. 

We shall then extend our analysis to study the CP violating
invariants for the Majorana fermions in the formalism of 
recursive parametrization of unitary matrices. One can study 
the CP phases in the neutrino mixing matrix and the 
Majorana phases through the recursive parametrization
of the unitary matrices. We shall extend these analyses and 
present explicit forms of the 
rephasing invariant quantities for a few examples when 
Majorana fermions are included.

\section{Majorana Fermions}

A Majorana fermion is the antiparticle of itself. It has two complex 
components or four real components. Two Majorana fermions may combine 
into a Dirac fermion, depending on the mass terms. The Lagrangian
describing a Majorana fermion can be given as
\begin{equation}
 {\cal L}_{M} = \overline{ \psi}_M ~ i \gamma_\mu \partial^\mu \psi_M 
 + m_M \overline \psi_M \psi^c_M 
\end{equation}
where $\bar \psi_M = \psi_M^\dagger \gamma^0 $ and we work in the Weyl 
representation, where the $\gamma$ matrices are defined as:
\begin{eqnarray}
    \gamma^\mu &=& \pmatrix{ 0 & \sigma^\mu \cr \bar \sigma^\mu & 0 }
    \hskip .15in {\rm with} ~~ \sigma^\mu = [I_2,\sigma_i]; ~
    \bar \sigma^\mu = [I_2,-\sigma_i] \nonumber
\end{eqnarray}
where $I_2$ is a $2 \times 2$ unit matrix and $\sigma^i$ are the
Pauli matrices. In this basis, $\gamma_5 = i \gamma^\circ \gamma^1
\gamma^2 \gamma^3 = {\rm diag}\pmatrix{ -I_2, & I_2 }$ is diagonal.
Defining the charge conjugation as
\begin{equation}
 \psi^c 
 = -i \gamma_2 \psi^\ast 
 = -i \gamma_2 \gamma_0 {\bar \psi}^{~T} \,,
 \label{cc}
\end{equation}
the Majorana condition that the Majorana particles are their
own antiparticles, can be written as
\begin{equation}
 { \psi^c_M }  = \lambda^* \psi_M \,.
 \label{Majph}
\end{equation}
here $\lambda$ is a complex phase contributing to CP
violation, and $|\lambda |^2 = 1$. 
Since the particles and antiparticles carry opposite quantum
numbers or charges under any symmetry group,
this condition implies violation of that quantum numbers
or the charges. So, charged leptons or quarks cannot be
Majorana particles. We shall thus work with the assumption
that neutrinos are Majorana particles, while all other 
charged fermions are Dirac particles. 

A Dirac fermion has eight real components (or equivalently, 
four complex components) and may be decomposed into
two Majorana fermions. For example, we can consider the 
left-handed ($\psi_L$) and right-handed ($\psi_R$) 
components of a Dirac fermion ($\psi_D$) as two Majorana 
fermions ($\psi_{M1}$ and $\psi_{M2}$) as defined below:
\begin{eqnarray}
 \psi_{M1} =  \psi_L + \lambda_1 ~{\psi^c}_R ~~~& {\rm and}& ~~~
 \psi_{M2} =  \psi_R +  \lambda_2 ~{\psi^c}_L  \nonumber \\
{\rm where} ~~~ ~~~ \psi_L = \frac{(1-\gamma_5)}{2} ~\psi_D ~~~; & & ~~~
 {\psi^c}_R = \frac{(1+\gamma_5)}{2} ~\psi^c_D~~~; \nonumber \\
 ~~~ ~~~ \psi_R = \frac{(1+\gamma_5)}{2} ~\psi_D ~~~ &{\rm and} & ~~~
 {\psi^c}_L = \frac{(1-\gamma_5)}{2} ~\psi^c_D ~\,.\nonumber 
\end{eqnarray}
and ${\psi^c}_R$ and ${\psi^c}_L$ are $CP$-conjugate
states of $\psi_L$ and $\psi_R$, respectively. 

The Majorana fermions $\psi_{M1}$ and $\psi_{M2}$ satisfy the 
Majorana conditions:
\begin{equation}
  { \psi^c_{M1} }  = \lambda_1^*~ \psi_{M1} \,~~~~{\rm and}~~~~
 { \psi^c_{M2} }  = \lambda_2^*~ \psi_{M2} \,.
\end{equation}
However, we can always remove an overall phase, so that 
the relative CP phases remain as independent phases. 

We can now express the Majorana fields $\psi_{M1}$ and
$\psi_{M2}$ in terms of complex two-component spinors
$\eta$ and $\chi$ as:
\begin{eqnarray}
    \psi_L = \frac{1 - \gamma_5}{2} \psi_M = \pmatrix{ \eta \cr 0 };~~~~~
    &\phantom{xx} &\psi_R = \frac{1 + \gamma_5}{2} \psi_M = \pmatrix{ 0 \cr 
\bar{ \chi}};
    \nonumber \\
    {\psi^c}_R = \pmatrix{ 0 \cr \bar{ \eta} };~~~~~
    &\phantom{xx} & {\psi^c}_L = \pmatrix{ \chi \cr 0}.
\end{eqnarray}
such that 
\begin{eqnarray}
    \psi_{M1} = \pmatrix{
    \eta \cr \lambda_1 ~\bar \eta};~~
    \psi_{M1}^c = \pmatrix{
    \lambda_1^\ast \eta \cr \bar \eta};~~
     \psi_{M2} = \pmatrix{
    \lambda_2~ \chi \cr \bar \chi};~~
    \psi_{M2}^c = \pmatrix{
    \chi \cr \lambda_2^\ast ~\bar \chi};~~
\end{eqnarray}
which satisfies the Majorana condition of equation (\ref{Majph}).

Any Majorana field can be expressed in terms of the
creation and the annihilation operators as
\begin{equation}
\psi_M (x) = \sum_{p,s} \sqrt{m_M \over 2 \epsilon } \left(
f_{ps} u_{ps} e^{-i p x} + \lambda^* f_{ps}^\dagger v_{ps} e^{i p x}
\right) \,,
\end{equation}
where the energy of the Majorana fermion is $\epsilon = \sqrt{p^2+m_M^2}$
and $m_M$ is the mass. The $u_{ps}$ and the
$v_{ps}$ spinor operators satisfy the equations of motion
\begin{equation}
 (\gamma_\mu p^\mu - m_M)~ u_p = 0; ~~~~~ (\gamma_\mu p^\mu + m_M) ~v_p = 0; 
\end{equation}
and $s$ is the spin.

To complete the discussion we shall define the density
matrix for the Majorana fermions as \cite{paschos,mavromatos}
\begin{equation}\label{Majrho}
    \rho^M (\psi) =  \psi^c  \left( \psi^c \right)^\dagger 
    = | \psi^c \rangle \langle \psi^c |
    \,,
\end{equation}
which satisfies the equation of motion
\begin{equation}
    \frac{d \rho^M}{dt} = - i~ (H \rho^M - \rho^M H^\dagger ) \,.
\end{equation}
This is a key ingredient in this analysis of Majorana fermions, which allows us
to define the smooth parametrized curves $C$ of unit vectors
in the Hilbert space $\H$ and the corresponding free geodesics. The Bargmann
invariants are then expressed in terms of these free geodesics and the geometric
phases can then be defined in terms of the Bargmann invariants. 

The density matrices $\rho_r$, corresponding to any unit
vector $\psi_r$ in the Hilbert space $\cal H$ of Majorana fermion
states, are images in the ray space $\cal R$ and any two
neighbouring density matrices (say, $\rho_{r-1}$ to $\rho_r$)
are connected by free geodesics in $\cal H$. In case of a 
Dirac fermion or Dirac neutrino, the density matrices or the images
of the unit vectors in $\H$ onto $\cal R$ is defined as
\begin{equation}\label{Dirrho}
    \rho^D (\psi ) = \psi ~{\psi}^\dagger =  | \psi \rangle  \langle \psi |
    \,.
\end{equation}
Thus the Dirac density matrix does not correspond to any violation of
charge or any conserved quantum number. The 
density matrix of the Majorana fermions would introduce the Majorana phases into
the density matrix giving rise to new sources of CP violation. 
The Majorana density matrix 
will provide us with additional Bargmann invariants corresponding
to the new Majorana CP phases $\lambda$ that is defined in
equation \ref{Majph}. The Majorana phases would disappear when
the neutrinos dont have Majorana masses. Furthermore, if there is
only one Majorana particle in any model, an overall
phase transformation can remove it. We shall elaborate on these
discussions in the next few sections.

\section{Bargmann Invariants and Geometric Phase for Majorana Fermions}

In this section we shall review the formalism of connecting the Bargmann
invariants with the geometric phases and extend the earlier results
by including the Majorana fermions, and hence, the Majorana phases. 
A connection between the Bargmann invariants and the geometric phases has
been established rigorously for the Dirac fermions \cite{BI-GP}. The basic
structure of this formalism relies on the cyclic adiabatic quantum-mechanical
evolution of the state vectors \cite{MukSim}. This has been further generalized to show
that the geometrical phases can be related to a family of quantum-mechanical
invariants which were proposed by Bargmann \cite{bargmann}. 
In this section we shall develop the connection between the geometrical phases
and the Bargmann invariants for the Majorana particles and discuss how these
studies can be extended to the CP violation in the lepton sector with 
Majorana neutrinos.

This analysis largely depends on the free geodesics in quantum-mechanical ray 
and 
Hilbert spaces, as the geometric phases vanish for these geodesics. It has been
demonstrated that the generalization of free geodesics to the so-called null
phase curves are more general in establishing a connection between the Bargmann 
invariants and the geometric phases. These null phase curves are a family of 
ray and Hilbert space curves, which includes the free geodesics and a large 
class
of other curves, and establish a general connection between the Bargmann 
invariants
and the geometric phases. However, for demonstrating the geometric phases of the
Majorana fermions and its connection with the Bargmann invariants, we shall 
restrict our discussions to free geodesics only, with the understanding that
these results are more general.

We start with a Hilbert space of some quantum system of both Dirac and 
Majorana particles $\H$, 
and construct the associated ray space $\R$ with the pure state density 
matrices.
The dual space of $\H$ will contain both particles and antiparticles for
the Majorama fermions, but for the Dirac fermions the dual space of $\H$ 
will contain only the particles. 
The density matrices for the Dirac fermions are defined in
equation \ref{Dirrho},
where $\psi_a$ represents a vector in the Hilbert space $\H$. 
The inner product of any two Dirac fermions would then be given by
\begin{equation} \label{DirInner}
 {\cal I}^D = \left( \psi_a (s),~ {\psi_b} (s) \right) = \langle \psi_a (s)
 | \psi_b (s) \rangle  \,.
\end{equation}

The key ingredient for studying a Majorana fermion in this formalism is
to enhance the ray space with the pure state density matrices for the
Majorana fermions defined by equation \ref{Majrho} in the previous
section. For the Majorana fermions, 
the charge conjugation of $\psi$ satisfy the Majorana condition of
equation \ref{Majph} and as a result differs from the Dirac 
fermions by the Majorana phase $\lambda$. 
Thus the ray space curves for the Majorana
fermions will differ from that of the Dirac fermions and this would modify the inner product of any two Hilbert space vectors. 

We can now write down an
inner product of two Majorana fermions $\psi_i$ and 
$\psi_j$ in $\H$ as
\begin{equation} \label{MajInner}
 {\cal I}^M = \left( \psi_i^c (s), {\psi^c_j} (s) \right) =
 \langle \psi_i^c(s) | \psi^c_j(s) \rangle  \,.
\end{equation}
For the Majorana fermions one can write both ${\cal I}^D$ and
${\cal I}^M$ type inner products, but for the Dirac fermions one can
only write ${\cal I}^D$ type inner products. 

If the geometric phase vanishes
for any connected part of the ray space curves, then the inner product of two
Hilbert space vectors along the lift of such ray space curves would also vanish.
This condition on the inner product of the Hilbert space vectors is also valid
for the free geodesics, which can also link the Bargmann invariants with the 
geometric phases. We shall now define the free geodesics in ray and Hilbert
spaces, in which the geometric phase vanishes and then demonstrate the 
connection between the Bargmann invariants and geometric phases. We shall
follow the formalism and notation of \cite{BI-GP,BI-GP1}. 

Any smooth parametrized curves $\C$ of unit vectors in $\H$ may then be
expressed as 
\begin{equation}
 \C = \{ \psi(s) \in \H~|~|| \psi(s) || =1, ~~ s_1 \leq s \leq s_2 \} \subset \H 
\,.
\end{equation}
The projection of the Hilbert space to the ray space $$\pi: \H \to \R. $$
will then provide us the image $\C_r$ in $\R$: $$ \pi[\C ] = \C_r \subset 
\R\,,$$
where the image $\C_r$ for the Majorana fermions follows from the definition
of the pure state density matrices for the Majorana fermions:
\begin{equation}
 \C_r = \{ \rho^M (s) = \psi^c(s) ~ {\psi^c}(s)^\dagger ~| ~ s_1 \leq s \leq s_2 
\} \,.
\end{equation}
Thus any curve of unit vectors $\C$ in the Hilbert space $\H$ 
($\C \subset \H$) is a lift of 
the image $\C_r$ in the ray space $\R$ ($\C_r \subset \R$). 
It is apparent that the end
points should satisfy the boundary condition that they are not orthonormal
\begin{equation}
 \left( \psi^c (s_1) , \psi^c (s_2) \right) \neq 0
\end{equation}
and  $\psi(s), \psi^c(s), \rho^M(s)$ and $\rho^D(s)$ are smooth curves, satisfying certain smoothness conditions \cite{BI-GP1}. 

We shall now consider the horizontal lift of the curve $\C^{(h)}_r \subset \R$,
such that the vectors $\psi^{c(h)}(s)$ along this lift satisfy 
\begin{equation}
 \left( \psi^{c(h)}(s), ~\frac {d}{ds} {\psi^c}^{(h)}(s) 
 \right)=0\,.
\end{equation}
This immediately implies vanishing of the dynamical phase for any curve $\C_r 
\subset \R$,
along the horizontal lift:
\begin{equation}
 \phi_{dyn} [ \C ] = {\rm Im} \int_{s_1}^{s_2} ds \left(
 \psi^{c(h)}(s), ~\frac{d}{ds} {\psi^c}^{(h)}(s) \right) = 0\,.
\end{equation}
The geometric phase is thus given by
\begin{equation} 
 \phi_g = {\rm arg} \left(~ \psi^c(s_1),~ {\psi^c}(s_2)~ \right) \ .
\end{equation}
One can then define the free geodesics in $\H$ and $\R$ and show that
the geometric phase vanishes along the free geodesics \cite{BI-GP}
\begin{equation}\label{freeG}
 \phi_g [{\rm free~ geodesics~ }\in~ \R ] =0 \,,
\end{equation}
and relate the geometric phases to the Bargmann variables. A more
general analysis utilizing the null phase curves can also establish
these relations \cite{BI-GP}, but for our purpose we shall directly
move to the final result. 

Bargmann invariants (BIs) were developed for the Dirac fermions, and the 
relationship with the geometric phase utilized the definition of the 
inner product (${\cal I}^D$) and the density matrices (${\rho^D}$) 
for the Dirac fermions \cite{bargmann}. To include the Majorana particles
and extend the applicability of the Bargmann invariants,
the Hilbert space $\H$ for the Dirac fermions is
enhanced to incorporate the antiparticles, and hence, the Majorana phases, 
as defined in equation \ref{Majph}.
The corresponding ray space is also modified by the definition \ref{Majrho}
of the density matrix (${\rho^M}$), and hence, the inner products involving the 
Majorana
fermions (${\cal I}^M$). 
Accordingly we can write down two types of the Bargmann invariants (BI)
for the Majorana fermions, one ($\Delta^D$) 
containing only ${\cal I}^D$ type
inner products, and the other ($\Delta^M$) 
containing both ${\cal I}^D$ and ${\cal I}^M$
type inner products. 

We first present a BI  with Dirac fermions. Since $\langle \psi_a| \psi_b 
\rangle = \langle
\psi_b | \psi_a \rangle^\ast $, all second order BIs are real and the 
corresponding 
geometric phase
vanishes. We thus present a third order BI with Dirac fermions:
\begin{eqnarray}
 \Delta_3^D (\psi_1, \psi_2, \psi_3) &=& \left(
 \psi_1, \psi_2 \right)
 \left(\psi_2, {\psi_3}\right) \left(\psi_3, {\psi_1} 
 \right) \nonumber \\
 &=& {\rm Tr} \left[\rho^D (\psi_1)~ \rho^D (\psi_2)~ \rho^D (\psi_3) \right] 
\nonumber \\
 &=& {\rm Tr} \left[ (\psi_1 {\psi_1}^\dagger )~
 (\psi_2 {\psi_2}^\dagger) ~(\psi_3 {\psi_3}^\dagger) \right] \ .
\end{eqnarray}
This is an example of a third order Bargmann invariant defined with
three mutually nonorthogonal vectors $\psi_i \in \H | i=1,2,3$ and
the ray space is defined by density matrices 
$\rho^D_i = \psi_i {\psi_i}^\dagger \in \R | ~i=1,2,3$. Any fourth or 
higher order BIs may be reduced to third order BIs. 
It is straightforward to generalize this definition to $m$-th order 
Bargmann invariants
\begin{eqnarray}
 \Delta_m^D (\psi_1, \psi_2, \cdots, \psi_m) &=& (\psi_1, {\psi}_2),
 (\psi_2, {\psi}_3), \cdots , (\psi_m, {\psi}_1) \nonumber \\
 &=& {\rm Tr} \left[ ~\rho^D (\psi_1)~ \rho^D (\psi_2) \cdots \rho^D 
 (\psi_m)~\right] \nonumber \\
 &=& {\rm Tr} ~\left[ \psi_1 {\psi_1}^\dagger
 ~\psi_2 {\psi_2}^\dagger ~\cdots ~\psi_m {\psi_m}^\dagger~ \right] \,.
\end{eqnarray}
From the properties of the free geodesics (equation \ref{freeG}), 
we can write down the relationship between the Bargmann 
invariants and the geometric phases for an $m$-vertex
closed loop (${\cal P}_m$) as,
\begin{equation}
 \phi_g [{\cal P}_m] = - {\rm arg}~\Delta_m^D 
 (\psi_1, \psi_2, \cdots, \psi_m)\,,
\end{equation}
\begin{eqnarray}
 {\rm where}~ &&  {\cal P}_m ~= ~m {\rm -vertex ~closed ~loop }~~ \in \R, 
\nonumber \\[.05in]
{\rm with} && \rho_1^D \to \rho^D_2 \to \rho^D_3 \to \ldots \to
\rho^D_m \to \rho^D_1 \ , \nonumber 
\end{eqnarray}
being connected by free geodesics. 

We now present an example of the third order BIs with Majorana fermions. These 
BIs are
possible only when the Hilbert space $\H$ incudes the antiparticles ($\psi^c$)
and both the density matrices  $\rho^D$ and $\rho^M$ appear in the definition of 
the BIs:
\begin{eqnarray}
 \Delta_3^M (\psi_1, \psi^c_2, \psi^c_3) &=& \left(
 \psi_1, \psi_2^c \right)
 \left(\psi^c_2, {\psi_3^c}\right) \left(\psi^c_3, {\psi_1} 
 \right) \nonumber \\
 &=& {\rm Tr} \left[ ~\rho^D (\psi_1)~ \rho^M (\psi_2^c) ~\rho^M (\psi^c_3)~ 
\right] \nonumber \\
 & =& {\rm Tr} \left[~ \psi_1 {\psi_1}^\dagger ~
 \psi_2 {\psi_2^c}^\dagger ~\psi_3 {\psi_3^c}^\dagger \right] \,.
\end{eqnarray}
both $\psi_i$ and $\psi^c_i$ enter in the definition of the BIs and the 
Majorana phases enter in the definition of the BIs through
the density matrices $\rho^M (\psi^c_i)$.
The consequences of the Majorana type BIs ($\Delta^M_i$) including the 
Majorana phases will become clear when we shall relate
them to the rephasing invariant measures of CP violation 
in the leptonic sector with Majorana neutrinos in the next section.

\section{Majorana neutrinos and BI}

The smallness of the neutrino mass can be explained in simple
extensions of the standard model without any fine tuning or introducing 
arbitrarily small parameters, by considering the neutrinos 
to be Majorana fermions. 
Any information about CP violation in the leptonic sector with Majorana
neutrinos are contained in the neutrino mass matrix and their charged
current interactions. Without loss of generality we can work in a basis, 
in which
the charged lepton mass matrix is diagonal, so that the complex phases
in the neutrino masses and mixing matrix will determine the CP violation
in any model. 

Some of the complex phases in the neutrino mass and mixing matrices can 
be removed by the rephasing of the neutrinos, so it is a general practice to 
construct rephasing invariant measures to study the CP violation. 
In this section we
shall demonstrate that these CP violating rephasing invariant measures 
are the Bargmann invariants with the Majorana neutrinos 
and are related to the geometric phases. In particular, we shall
emphasize on the lepton number violating CP violating measures, which
are the new Bargmann invariants with the Majorana fermions,

We begin with the charged current interactions of the neutrinos
with the charged leptons and the neutrino mass matrix for the 
Majorana neutrinos:
\begin{eqnarray}
 {\cal L}_{CC} &=& -\frac{g}{\sqrt{2}} ~\sum_{\alpha=e,\mu,\tau}  
 \bar{\nu}_{\alpha L} \gamma^\rho
 l^-_{\alpha L} W^+_\rho \ , \nonumber \\
 {\cal L}_{mass} &=& \nu_{iL}^T~ C^{-1} ~\nu_{iL} = m_i~ 
 \overline{{\nu_{iL}}^c} ~\nu_{iL} \ ,
\end{eqnarray}
where $\nu_{iL},~i=1,2,3$ are the three left-handed Majorana neutrinos. 
The corresponding neutrino mass  matrix is diagonal with eigenvalues $m_i$ and  
$l_{\alpha L}, ~\alpha=e,~\mu,~\tau$ are the weak charged lepton 
eigenstates, which are the states with diagonal charged 
lepton mass matrix. The Majorana neutrinos in this
basis $\nu_{\alpha L}$ are related to the physical neutrinos 
$\nu_{iL}$ by a unitary transformation \cite{Bilenky_Petcov} given by
\begin{equation} \label{mixing}
\nu_{i L} = \sum_{\alpha=e,\mu,\tau} \left( U_{\alpha i}^*~ \nu_{\alpha L} 
 + {\lambda_i} 
 ~U_{\alpha i} ~{\nu^c}_{R \alpha} \right) \ .
\end{equation}

$U$ is the neutrino mixing matrix. Since the 
right-handed fermions are blind to the $SU(2)_L$ interactions, we
shall be working with only the left-handed Majorana neutrions. 
So we shall drop the index $L$.

The mixing matrix $U_{\alpha i}$ that relates the weak neutrino
eigenstates $\nu_\alpha $ to the physical neutrino eigenstates
$\nu_i$:
\begin{equation} \label{mixmat}
%
 |~\nu_{i} \rangle = \sum_{\alpha=e,\mu,\tau} ~\left(~ U^*_{\alpha i}~ 
 |~ \nu_{\alpha} \rangle 
 + {\lambda_i} 
 ~U_{\alpha i} ~|~{\nu^c_\alpha } \rangle ~\right)  \,.
\end{equation}
appears in the charged
current interactions of the physical neutrinos $\nu_i$ with the 
physical charged leptons $l_{\alpha}$:
$$
{\cal L}_{CC} 
 = -\frac{g}{\sqrt{2}} ~\sum_{\alpha=e,\mu,\tau}  
 \bar{\nu}_{i }~ {(U^\dagger)}_{i \alpha} ~\gamma^\rho~
 l^-_{\alpha }~ W^+_\rho + H.c.
 $$
and it relates the neutrino 
mass matrix in this weak interaction basis to the physical neutrino mass matrix 
(diagonal) by
\begin{equation} \label{diagM}
(U^T)_{i \alpha }~ M_{\alpha \beta} ~ U_{\beta j} = \lambda^\ast_i ~ M_{ij}^{\rm 
diag}\,.
\end{equation}
$\lambda_i$ are the Majorana phases defined by equation \ref{Majph},
so that the physical Majorana neutrinos $\nu_i$ also satisfy the 
Majorana condition 
$$ \nu_i^c ={\lambda_i^*} ~\nu_i.$$
Any complex phases in the mass matrix $M_{\alpha \beta}$ may be tansferred to 
the mixing matrix $U_{\alpha i}$ by the rephasing of the physical neutrinos
and the weak basis states of neutrinos, but the Majorana phases may not be removed 
independently. 
Rephasing the gauge basis and the mass basis of the neutrinos
\begin{equation}
 \nu_{i} \to e^{ - i \delta_i} \nu_{i} ~~~ {\rm and} ~~~ 
\nu_\alpha \to e^{ - i \eta_\alpha} \nu_\alpha \, ,
\end{equation}
would then imply rephasing of the mixing matrix and the Majorana phase
matrix as:
\begin{equation}
 U_{\alpha i} \to e^{-i (\eta_\alpha - \delta_i)} U_{\alpha i}~, ~~~~
 \lambda_i \to e^{- 2 i \delta_i} \lambda_i
 ~~~ {\rm and} ~~~ 
 \tilde \lambda_i \to e^{i \delta_i} \tilde \lambda_i \,.
\end{equation}
where we defined $\lambda^\ast = \tilde \lambda^2$.

Both Dirac and Majorana neutrinos can have CP violation coming from
the phases in the mixing matrix U. The simplest 
rephasing invariant combination with the mixing matrix can be
defined as \cite{np}
\begin{eqnarray} \label{rephasing}
T^D_{\alpha i \beta j} &=& U_{\alpha i } U_{\beta j} U^\ast_{\alpha j} 
U^\ast_{\beta i} 
\end{eqnarray}
While the rephasing invariant measure $T^D_{\alpha i \beta j}$ contains all the 
CP violating
phases in the mixing matrix, it does not include the Majorana phases,
and hence, any CP violation in a lepton number violating interaction
will not have any contribution from this measure $T^D_{\alpha i \beta j}$.

The simplest rephasing invariant measure containing the Majorana phases
consists of two mixing matrix and the Majorana phase matrices \cite{np}
\begin{eqnarray} 
 s^M_{\alpha ij} &=& U_{\alpha i} U^\ast_{\alpha j} \tilde \lambda_{i}^\ast
 \tilde \lambda_j ~\,\,.
\end{eqnarray}
Although this rephasing invariant measure contains the Majorana phases,
it may not appear in the probability or cross-section of 
any lepton number violating physical processes. The rephasing invariant
measure that contains the Majorana phase and also enter in the 
physical processes may be defined as \cite{kiku}
\begin{eqnarray} 
T^M_{\alpha i \beta j} = U_{\alpha i} U^*_{\beta j} U^*_{\alpha j} U_{\beta i} \lambda_i \lambda^*_j
\end{eqnarray}
We shall now demonstrate that the measures $T^D_{\alpha i \beta j}$ and $T^M_{\alpha i \beta j}$ may be defined as Dirac and Majorana type Bargmann invariants and can be viewed as geometric phases. 

We start with the state vectors $|\nu_i \rangle $,~ $|\nu_\alpha
\rangle $ and $|\nu_\alpha^c
\rangle $  in the Hilbert space $\H$. The state vector $|\nu_i \rangle $
satisfy the Majorana condition of equation \ref{Majph}, that is,
$\nu_i^c = \lambda _i \nu_i$ and can be expressed in terms of the
state vectors $| \nu_\alpha \rangle $ and $|\nu_\alpha^c \rangle $ 
as given by equation \ref{mixmat}. We can then utilize the 
orthogonality conditions of the state vectors $|\nu_i \rangle $,~ $|\nu_\alpha
\rangle $ and $|\nu_\alpha^c \rangle $ $$ \langle \nu_a | \nu_b \rangle 
= (\nu_a, \nu_b) = \delta_{ab} $$
where $\nu_{a,b} \equiv \nu_{\alpha, \beta} \,, 
~~\nu_{i,j} ~~ {\rm or}~~ \nu^c_{\alpha, \beta}$, and using
equation \ref{mixmat}, express the 
inner products of the non-orthogonal state vectors as
\begin{eqnarray}
 (\nu_i,\nu_\alpha ) &=& U_{ \alpha i} \nonumber \\
 (\nu_i,\nu^c_\alpha ) &=& \lambda_i^\ast U^\ast_{ \alpha i}
\end{eqnarray}
We now construct Bargmann invariants without Majorana phases
and with only the mixing matrix, which is
\begin{eqnarray}
 \Delta^D_4 (\nu_i, \nu_\alpha, \nu_j, \nu_\beta ) &=&  (\nu_i, \nu_\alpha ) ( 
\nu_\alpha, \nu_j) (\nu_j, \nu_\beta)
 (\nu_\beta,  \nu_i)  \nonumber \\
 &=& {\rm Tr}~\left[ \rho^D( \nu_i) \rho^D(\nu_\alpha) \rho^D(\nu_j)
 \rho^D(\nu_\beta) \right] \nonumber \\
 &=& {\rm Tr} ~ \left[ (\nu_i \nu_i^\dagger ) (\nu_\alpha \nu_\alpha^\dagger )
 (\nu_j \nu_j^\dagger ) (\nu_\beta \nu_\beta^\dagger ) \right] \nonumber \\
 &=& U_{ \alpha i} U^\ast_{\alpha j } U_{ \beta j} U^\ast_{ \beta i}
 \nonumber \\
 &=& T^D_{\alpha i \beta j}
\end{eqnarray}
Similarly we can construct the Bargmann invariants with Majorana fermions,
and hence, Majorana phases $\lambda_i$. It should contain both $\nu_\alpha $ and
$\nu_\alpha^c$ and the simplest one is given by
\begin{eqnarray}
\Delta^M_4 (\nu_i, \nu_\alpha,  \nu_j, \nu_\beta^c ) &=&  (\nu_i ,\nu_\alpha) (\nu_\alpha , \nu_j) (\nu_j , \nu_\beta^c) (\nu_\beta^c , \nu_i) \nonumber \\
 &=& {\rm Tr}~\left[ \rho^D(\nu_i) \rho^D(\nu_\alpha) \rho^D(\nu_j) \rho^M(\nu_\beta^c) \right] \nonumber \\
 &=& {\rm Tr} ~ \left[ (\nu_\alpha \nu_\alpha^\dagger ) (\nu_i \nu_i^\dagger )
 (\nu_j \nu_j^\dagger ) (\nu^c_\beta {\nu_\beta^c}^\dagger ) \right] \nonumber 
\\
 &=& U_{\alpha i} U^*_{\alpha j} \lambda^*_j U^*_{\beta j} \lambda_i U_{\beta i} \nonumber \\
 &=& T^M_{\alpha i \beta j}
\end{eqnarray}
This is the rephasing invariant measure of CP violation with Majorana fermions.
It is clear from the form of $s_{\alpha ij}$ that it can enter in any
Bargmann invariants, but it cannot be a Bargmann invariant because it
is not closed. The Bargmann invariants we constructed $\Delta^D_4$ and
$\Delta^M_4$, are the conventional rephasing invariant measures
$T^D_{\alpha i \beta j}$ and $T^M_{\alpha i \beta j}$, respectively.
Moreover, the phases in $s_{\alpha ij}$ are related to these BIs $T^D_4$
and $T^M_4$ 
\begin{equation}
 T^D_{\alpha i \beta j} = s_{\alpha ij} ~ s^\ast_{\beta ij} ~~~~
 {\rm and} ~~~~ 
 T^M_{\alpha i \beta j} = s_{\alpha ij} ~ s_{\beta ij} 
\end{equation}

The Bargmann variables, $\Delta^D_4$ and
$\Delta^M_4$, are defined on the ray space $\R$ and the points
$\rho(\nu_i), \rho(\nu_\alpha), \rho(\nu_\alpha^c)$ on $\R$ are
non-orthogonal and pair wise linearly independent. These points 
are connected by geodesics, which form a closed loop in $\R$, 
representing a cyclic evolution in the state space. This establishes
that the Bargmann invariants, and hence, the rephasing invariant 
measures, gives us the geometric phases
$$ \phi_g = - {\rm arg} (\Delta_4)$$
The Majorana nature of the neutrinos implies lepton number violation. So, 
$\Delta_4^M$
is the rephasing invariant measure of CP violation \cite{kiku} that enters in the lepton 
number violating CP
violating interactions like the neutrinoless double beta decays
or $W^-~ W^- \to e^-~e^-$. If we extend this analysis to include
the right-handed neutrinos, then a similar CP violating measure with the 
right-handed neutrinos would appear 
in the lepton number violating CP asymmetry as in the models of leptogenesis \cite{ODon}.

\section{Recursive Parametrization and rephasing invariants of unitary matrices 
}

In the previous section we constructed the Bargmann invariants with
Majorana neutrinos and demonstrated that the CP violation coming from the 
Majorana phase $\lambda_i$ and also the unitary mixing matrix $U_{i \alpha}$
can be presented in the form of rephasing invariant measures or 
equivalently as BIs. This establishes that the CP violating phases
are geometric phases in the leptonic sector. In this section we shall
extend this analysis to study the unitary mixing matrices
in the framework of recursive parametrization
and demonstrate how to include the Majorana phases in this formalism. 

We shall first define the recursive parametrization of any
unitary matrix, keeping in mind that we shall be applying this
formalism to discuss the neutrino mixing matrix defined by
equations \ref{mixmat} and \ref{diagM}. Then we shall generalize the 
formalism to incorporate the Majorana phases, defined by equation \ref{diagM}.
For the unitary matrix without including the Majorana phases, we
shall use the notation and conventions of reference \cite{4}. 

Any $n \times n$ matrix $A_n \in U(n)$ can be uniquely decomposed 
[4] into $n$ block matrices given by
\begin{equation}
A_n = A_n(\zeta) A_{n-1}(\eta) A_{n-2}(\xi) ... 
A_4(\gamma)A_3(\beta)A_2(\alpha)A_1(\chi)
\label{eqn:decomp}
\end{equation}
The elements $a_{jk}$ of $A_n(\zeta)$ matrix can be constructed in the 
following 
way
\begin{eqnarray}
	a_{jn} &= & \zeta_j ; \hspace{0.5cm} j = 1,2,3 ... n \nonumber \\
	a_{j\ j-1} &=& \frac{\rho_{j-1}}{\rho_j}; \hspace{0.5cm} j = 2,3 ... n 
	\nonumber \\
	a_{jk} &=&  - \frac{\zeta_j \ {\zeta^*}_{k+1}}{\rho_k \ \rho_{k+1}}; 
\hspace{0.5cm} j \leq k \leq n-1 \nonumber \\
 	a_{jk} &=& 0; \hspace{0.5cm} \forall \hspace{0.5cm} j \geq k+2 
\label{eqn:cond}
\end{eqnarray}
$\zeta$ is the unit vector forming the basis of the $n \times n$ space implying 
the components $\zeta_i$ to obey $\sum\limits_{i=1}^{n}\zeta_i^2 = 1$ and 
$\rho_j = \sqrt{\sum\limits_{i=1}^{j} |\zeta_i|^2}$ . Any $A_m \in U(m) \ 
(m<n)$ 
is an unitary matrix with diagonal elements equal to $1$ and all other elements 
equals 
to zero for trivial rows and columns. This method can also be used to construct 
$SU(n)$ matrices by multiplying the first column of the obtained $U(n)$ matrix 
with $(-1)^{n-1} \frac{{\zeta^*}_1}{\zeta_1}$ assuming $\zeta_1 \neq 0 $. 

We shall now proceed to construct the rephasing invariants in this formalism. 
Once we have the required $U(n)$ matrix, we consider the freedom to
rephase the various states by multiplying $U(n)$ matrix with
the diagonal phase matrices 
\begin{eqnarray}
  U^\prime &=& D(\theta) ~ U ~ D(\theta^\prime)  \nonumber \\
 \mathrm{where}~~~~ D(\theta) &=& \rm{diag}( 
e^{i\theta_1}, e^{i\theta_2}, e^{i\theta_3}, ... , e^{i\theta_n} ) \,.
\end{eqnarray}
For the appropriate choice of the phases $\theta$ and $\theta^\prime$, 
one can get explicit form of the rephasing invariants in this formalism.
Before we consider explicit construction of such invariants, we shall
demonstrate how these constructions differ for the Dirac and Majorana
cases, so that we can demonstrate the Majorana phases for the 
groups $SU(n)$ for $n=2,3,4$ that we shall study.

So far we have considered
the Hilbert space $\H$ with only the Dirac neutrinos. The mass term for a Dirac
neutrino can be written as
$$ {\cal L}_{mass}^D = m_D ~\bar \nu_L ~ M_\nu^D ~\nu_R $$ which can be 
diagonalized
by a bi-unitary transformation
\begin{equation}
 U_L^\dagger ~ M_\nu^D ~ U_R = M_\nu^{diag} \,.
\end{equation}
where $U_L$ and $U_R$ diagonalizes the matrices $M_\nu^D ~{M_\nu^D}^\dagger $
and ${M_\nu^D}^\dagger ~{M_\nu^D} $, respectively. 

If we make any phase transformation to the left-handed and the 
right-handed neutrinos by the matrices $D(\theta)$ and $D(\theta^{\prime 
\prime})$
then 
\begin{eqnarray}
M^{diag}_\nu &=& U_L^\dagger~ M_\nu^D ~U_R \nonumber \\
&=& U_L^\dagger ~ D^\dagger(\theta)~
M_\nu^D~ D(\theta^{\prime \prime})~ U_R \nonumber \\ 
&=&D^\dagger (\theta^\prime)~ U_L^\dagger~ D^\dagger (\theta )~
M_\nu^D ~D(\theta^{\prime \prime})~ U_R~ D (\theta^\prime) \nonumber \\
&=& {U_L^{\prime}}^\dagger~ M^D_\nu {U_R^\prime}  \,. 
\end{eqnarray}
Thus the phase transformation of the left-handed neutrinos
can be represented by the transformation of $U_L$ as
\begin{equation}
 U_L \to {U_L^\prime} = D(\theta)~ U_L~ D(\theta^\prime)   \,.
 \label{eqn:dirac_rephase}
\end{equation}
Thus for the Dirac fermions, we have the freedom to make two sets
of rephasing with the parameters $\theta$ and $\theta^\prime$.

Since the right handed neutrinos $\nu_R$ do not enter the 
$SU(2)_L \times U(1)_Y$ charged current
 interactions, their transformation (rephasing of the 
matrix $U_R$ by $D$) will not affect our analysis. 

In case of 
Majorana neutrinos, the mass term may be written as 
\begin{equation}
 {\cal L}_{mass}^M = m_M ~ \overline{ {\nu_L}^c} ~ M_\nu^M ~ \nu_L \,.
\end{equation}
This mass matrix $M_\nu^M$ is symmetric and may be diagonalized by
only one matrix $U$, and hence, 
the same unitary matrix $U$ will diagonalize
the neutrino mass matrix, and hence,
\begin{eqnarray}
  \lambda_i M_\nu^{diag} &=&  U^T ~M^M_\nu ~U \nonumber \\
 &=& U^T ~ D(\theta)~
M_\nu^M~ D(\theta^{})~ U
\end{eqnarray}
Thus rephasing of the Majorana fermions may be represented by
\begin{equation}
 U \to {U^\prime} = D(\theta)~ U \ .
 \label{eqn:majorana_rephase}
\end{equation}
Given the prescription for the rephasing of Dirac neutrinos
by equation \ref{eqn:dirac_rephase} and for the rephasing of the Majorana
neutrinos by equation \ref{eqn:majorana_rephase}, we can now 
explicitly construct the rephasing invariants in this recursive
parametrization formalism for both the Dirac and Majorana fermions,
as demonstrated below for the groups $SU(n)$, $n=2,3,4$.

\subsection{Parametrization of $SU(2)$}
The $SU(2)$ matrix obtained in the formulation of recursive 
parametrization is given by
\[
U = A_2(\alpha) = 
\pmatrix{
{\alpha_2}^*     &     \alpha_1 \cr
-{\alpha_1}^*    &     \alpha_2
}
\]
Choosing the diagonal phase matrix as $D(\theta) = \mathrm{diag} 
(e^{i(\theta_1)}, 
e^{i(-\theta_1)})$ and rephasing according to equation \ref{eqn:dirac_rephase}, 
for the Dirac neutrinos, it is clear that the matrix remains unchanged i.e. 
$U^\prime = U$. Whereas, rephasing for Majorana neutrinos according to equation 
\ref{eqn:majorana_rephase} gives
\[
U^\prime = D(\theta) U = 
\pmatrix{
{\alpha_2}^* e^{i\theta_1}    &     \alpha_1 e^{i\theta_1} \cr
-{\alpha_1}^* e^{-i\theta_1}   &     \alpha_2 e^{-i\theta_1}
}.
\]
The action of rephasing changes the elements of the matrix as
\begin{eqnarray*}
\alpha_1 \rightarrow \alpha_1^\prime &=& \alpha_1 e^{i\theta_1} \ ,\\
\alpha_2 \rightarrow \alpha_2^\prime &=& \alpha_2 e^{-i\theta_1} \ ;
\end{eqnarray*}
which can be written in general as $\zeta_j \rightarrow \zeta_j^\prime 
= \zeta_j e^{i n_1 \theta_1}$ and can be represented in a tabular form in table 
\ref{table:SU(2)}. 
\begin{table}[ht]
\centering
\renewcommand{\arraystretch}{1.5}
\begin{tabular}{|c|c|}
\hline 
{$\zeta_j \rightarrow \zeta_j^\prime$} & $\zeta_j e^{i n_1 
\theta_1}$\tabularnewline
\cline{2-2} 
 & $n_1$\tabularnewline
\hline 
\hline
$\alpha_1 \rightarrow \alpha_1^\prime$ & $+1$ \tabularnewline
\hline 
$\alpha_2 \rightarrow \alpha_2^\prime$ & $-1$ \tabularnewline
\hline 
\end{tabular} .
\caption{Changed components of unit vectors after rephasing $SU(2)$}
\label{table:SU(2)}
\renewcommand{\arraystretch}{1}
\end{table}

The number of independent elements of the unit vector (in this case 
$\vec{\alpha}$) decreases by one after rephasing due to the constraint 
$\alpha_2^\prime = {\alpha_1^\prime}^*$, leaving only one rephasing invariant 
quantity given by $(\alpha_1 \alpha_2)$. It is important to note that no such 
quantities are found for Dirac type rephasing which reflect the fact that only 
two generations of quark cannot produce CP violating pure Dirac phase whereas 
it is sufficient to produce CP violating majorana phase in the lepton sector
with two generations of Majorana neutrinos.

\subsection{Recursive parametrization of $SU(3)$}

The recursive parametrization scheme \ref{eqn:cond} can be easily extended to 
obtain $SU(3)$ matrices in the form given by
\begin{eqnarray*}
U &=& A_3(\beta)A_2(\alpha) \\
&=& \pmatrix{\displaystyle 
\frac{-\beta_2^* \alpha_2^* + \beta_3^* \beta_1 \alpha_1}{\sigma_2}	&	
-\displaystyle \frac{\beta_2^* \alpha_1 + \beta_3^* \beta_1 \alpha_2}{\sigma_2} 	
& 	
\beta_1  \cr
\displaystyle \frac{\beta_1^* \alpha_2^* + \beta_3^* \beta_2 
\alpha_1^*}{\sigma_2}	& 	
\displaystyle \frac{\beta_1^* \alpha_1 - \beta_3^* \beta_2 \alpha_2}{\sigma_2} 	
& 	
\beta_2  \cr
-\sigma_2 \alpha_1^* &	\sigma_2 \alpha_2 & \beta_3 } , \sigma_2 = 
\sqrt{|\beta_1|^2 + |\beta_2|^2} \ .
\end{eqnarray*}

Choosing the diagonal phase matrix as $D(\theta) = \mathrm{diag} (e^{i(\theta_1 
+ 
\theta_2)}, e^{i(-\theta_1 + \theta_2)}, e^{i(-2\theta_2)})$ and after rephasing 
by Dirac type i.e. $U^\prime \rightarrow U = D(\theta) U D(\theta^\prime) $, we 
get the changed components of unit vectors given in the table \ref{table:SU(3)}
\begin{table}[ht!]
\renewcommand{\arraystretch}{1.5}
\centering
\begin{tabular}{|c|c|c|c|c|}
\hline 
{$\zeta_j \rightarrow \zeta_j^\prime = $} & \multicolumn{4}{c|}{$\zeta_j e^{i 
n_1 \theta_1 + i n_2 \theta_2 + i n_1^\prime \theta_1^\prime + i n_2^\prime 
\theta_2^\prime}$}\tabularnewline
\cline{2-5}
 & $n_1$ & $n_2$ & $n_1^\prime$ & $n_2^\prime$ \tabularnewline
\hline  \hline
$\alpha_1 \rightarrow \alpha_1^\prime = $  & $0$ & $+2$ & $-1$ & 
$-1$\tabularnewline 
\hline 
$\alpha_2 \rightarrow \alpha_2^\prime = $  & $0$ & $-2$ & $-1$ & 
$+1$\tabularnewline
\hline 
$\beta_1 \rightarrow \beta_1^\prime = $ & $+1$ & $+1$ & $0$ & 
$-2$\tabularnewline
\hline 
$\beta_2 \rightarrow \beta_2^\prime = $ & $-1$ & $+1$ & $0$ & 
$-2$\tabularnewline
\hline 
$\beta_3 \rightarrow \beta_3^\prime = $ & $0$ & $-2$ & $0$ & 
$-2$\tabularnewline
\hline 
\end{tabular}
\renewcommand{\arraystretch}{1}
\caption{Changed components of unit vectors after rephasing for $SU(3)$}
\label{table:SU(3)}
\end{table}

It is interesting to note that the only quantity which is linear in every 
component of unit vectors (in this case $\vec{\alpha}$ and $\vec{\beta}$) and 
remains invariant after rephasing is given by $ \alpha_1 \alpha_2^* \beta_1^* 
\beta_2^* \beta_3 $. This fact reflects that only one pure Dirac phase can occur 
in the mixing matrix for CP violation with three generations.

However, using Majorana type rephasing for the mixing matrix (i.e $U^\prime 
\rightarrow U = D(\theta) U $) leads to more invariant quantities. Also the 
number of independent components of unit vectors reduces to three after 
rephasing due to the constraint $\alpha_2^\prime = {\alpha_1^\prime}^* = 
\beta_3^\prime$. The two smallest forms of rephasing invariants,each of which is 
linear in each component of unit vectors in this case are $(\alpha_1 \alpha_2)$ 
and $(\beta_1 \beta_2 \beta_3)$.

\subsection{Recursive parametrization of $SU(4)$}

The procedure can be extended easily to four generations of leptons to get 
$SU(4)$ mixing matrix in the form of $ U = A_4(\gamma) A_3(\beta) A_2(\alpha)$. 
We choose the diagonal phase matrix for Dirac type rephasing to be $ D(\theta) = 
\mathrm{diag}(e^{i(\theta_1 + \theta_2 + \theta_3)}, \\
e^{i(-\theta_1 + \theta_2 + \theta_3)}, e^{i(-2 \theta_2 + \theta_3)}, e^{i(-3 
\theta_3)}) $. The set of changes in the components of the unit vectors after 
rephasing in the form $U \rightarrow U^\prime  = D(\theta) U D(\theta^\prime) $ 
is given in table \ref{table:SU(4)}.

\begin{table}[H]
\renewcommand{\arraystretch}{1.5}
\centering
\begin{tabular}{|c|c|c|c|c|c|c|c|}
\hline 
$\zeta_j \rightarrow \zeta_j^\prime = $ & \multicolumn{6}{c|}{$\zeta_j e^{i n_1 
\theta_1 + i n_2 \theta_2 + i n_3 \theta_3 + i n_1^\prime \theta_1^\prime + i 
n_2^\prime \theta_2^\prime + i n_3^\prime \theta_3^\prime}$}\tabularnewline
\cline{2-7} 
 & $n_1$ & $n_2$ & $n_3$ & $n_1^\prime$ & $n_2^\prime$ & 
$n_3^\prime$\tabularnewline
\hline 
\hline
$\alpha_1 \rightarrow \alpha_1^\prime = $ & $0$ & $0$ & $+3$ & $-1$ 
& $-1$ & $-1$\tabularnewline
\hline 
$\alpha_2 \rightarrow \alpha_2^\prime = $ & $0$ & $0$ & $-3$ & $-1$ 
& $+1$ & $+1$\tabularnewline
\hline 
$\beta_1 \rightarrow \beta_1^\prime = $ & $0$ & $+2$ & $+2$ & $0$ & 
$-2$ & $-2$\tabularnewline
\hline 
$\beta_2 \rightarrow \beta_2^\prime = $ & $0$ & $-2$ & $+1$ & $0$ & 
$-2$ & $+1$\tabularnewline
\hline 
$\beta_3 \rightarrow \beta_3^\prime = $ & $0$ & $0$ & $-3$ & $0$ & 
$-2$ & $+1$\tabularnewline
\hline 
$\gamma_1 \rightarrow \gamma_1^\prime = $ & $+1$ & $+1$ & $+1$ & 
$0$ & $0$ & $-3$\tabularnewline
\hline 
$\gamma_2 \rightarrow \gamma_2^\prime = $ & $-1$ & $+1$ & $+1$ & 
$0$ & $0$ & $-3$\tabularnewline
\hline 
$\gamma_3 \rightarrow \gamma_3^\prime = $ & $0$ & $-2$ & $+1$ & $0$ 
& $0$ & $-3$\tabularnewline
\hline 
$\gamma_4 \rightarrow \gamma_4^\prime = $ & $0$ & $0$ & $-3$ & $0$ 
& $0$ & $-3$\tabularnewline
\hline 
\end{tabular}
\caption{Changed components of unit vectors after rephasing for $SU(4)$}
\label{table:SU(4)}
\renewcommand{\arraystretch}{1.5}
\end{table}

There are three pure Dirac phases appearing in the mixing matrix for four 
generations and are related to the invariant quantities which are given as 
$(\alpha_1 \alpha_2^* \beta_1^* \beta_2^* \beta_3)$, $(\beta_2 \beta_3^* 
\gamma_3^* \gamma_4)$ and $(\beta_1 \beta_2^* \gamma_1^* \gamma_2^* \gamma_3)$. 

On the other hand, Majorana type rephasing constraints the number of independent 
components of unit vectors by ${\alpha_1^\prime}^* = \alpha_2^\prime = 
\beta_3^\prime = 
\gamma_4^\prime$ and $\beta_2^\prime = \gamma_3^\prime$. The three invariant 
quantities of smallest forms are given by $(\alpha_1 \alpha_2)$, $(\beta_1 
\beta_2 \beta_3)$ and $ (\gamma_1 \gamma_2 \gamma_3 \gamma_4)$.

It is quite evident that rephasing invariant quantities are structurally 
different due to the additional constraints appearing only in Majorana type 
rephasing. All the invariant quantities are of lower order and linear in every 
component of unit vectors required to construct the unitary matrix recursively. 
Higher order invaraints can be constructed out of these lower order invariants. 
For $n$ generations of leptons \cite{np}, the number of CP violating pure Dirac 
phases in the mixing matrix is $(n-1)(n-2)/2$, whereas the number of pure 
Majorana phases is $(n-1)$ with a total number of $n(n-1)/2$ phases. It is 
important to note that the number of Dirac and Majorana phases are exactly equal 
to the number of rephasing invariants for the Dirac type and Majorana type 
rephasing respectively. The parametrization fails if the first component of 
$\vec{\zeta}$ (in equation \ref{eqn:decomp}) vanishes. In such a case 
\cite{second_type_par}, a different recursive approach produces different form 
of unitary matrix which upon rephasing changes the components of the unit vector 
in a different way, however, the number of rephasing invariants and the forms 
remain unchanged.


\section{Summary}

Bargmann invariants have been shown to connect the rephasing invariant quantities 
with the geometric phase for the Dirac fermions. We extend this analysis to include
the Majorana fermions and show how to relate the Majorana phases with the Bargmann
invariants, by defining proper quantum mechanical ray and Hilbert spaces for the 
Majorana fermions. As an explicit example, we included Majorana neutrinos in the leptonic sector
and constructed the Bargmann invariants for both the Dirac and the Majorana neutrinos.
This allows us to interpret the CP violating phases of Dirac and Majorana fermions
to a geometric phase. We then explained how to incorporate Majorana phases in
a recursive parametrization of unitary matrices with explicit examples of SU(n), n = 2,3,4.

\end{document}